\documentclass[journal]{IEEEtran}
\usepackage{graphicx}
\ifCLASSINFOpdf
  
\else

\fi

\usepackage[english]{babel}
\usepackage[dvipsnames,svgnames,x11names]{xcolor}
\usepackage[markup=underlined]{changes}

\usepackage[cmex10]{amsmath}

\begin{document}

\title{
  Widlar Current Mirror Design Using BJT-Memristor Circuits\\
}

\author{Amanzhol~Daribay and Irina Dolzhikova,\\
   
\IEEEauthorblockA{Electrical and Computer Engineering Department, Nazarbayev University, Astana, Kazakhstan\\
amanzhol.daribay@nu.edu.kz, ifedorova@nu.edu.kz}
}

\maketitle

\IEEEpeerreviewmaketitle
\begin{abstract}
This paper presents a description of basic current mirror (CM), Widlar current mirror, fourth circuit element (memristor) and an analysis of Widlar Configuration with integrated memristor.  The analysis has been performed by comparing a modified configuration with a simple circuit of Widlar CM. The focus of analysis were a power dissipation, a Total Harmonic Distortion and a chip-surface. The results has shown that a presence of memristor in the Widlar CM decreases the chip-surface area and the deviation of the signal in the circuit from a fundamental frequency. Although the analysis of power dissipation has also been conducted, there is no definite conclusion about the power losses in the circuit because of the memristor model. 
\end{abstract}

\begin{IEEEkeywords}
current mirror, Widlar current source, bjt-memristor circuit, power analysis, noise analysis, total harmonic distortion.
\end{IEEEkeywords}

\section{Introduction}
\subsection{Basic Current Mirror}

\IEEEPARstart {} \indent
Basically, current mirrors (CMs) are used to mirror a reference current multiple times from one designated source into another consuming circuits. For example, as it is shown in figure below, current from Ideal Reference Generator has been copied two times, specifically, into $I_1$ and $I_2$. 
\begin{figure}  [!ht] 
\begin{center}  
\includegraphics[scale=0.3]{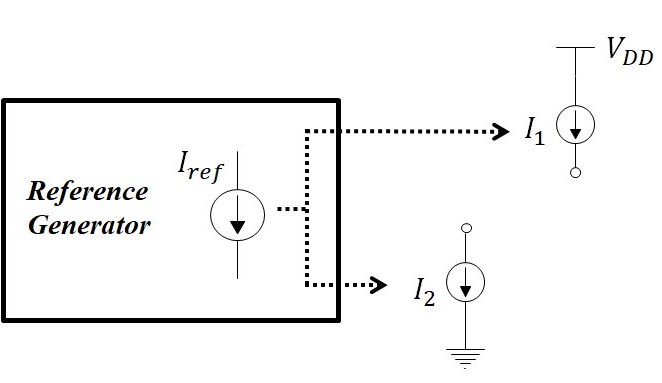} 
\caption{\small \sl Representation of CMs working principle}
\label{noise_basic} 
\end{center}     
\end{figure}
This kind of and many other current mirrors are considered to be extremely useful in integrated circuits, multipliers, differential amplifiers, operational amplifiers and etc. 

A basic current mirror is a circuit consisting from two bipolar junction transistors (BJTs) that have common base connection and a simple resistor at the input. To construct the circuit of Basic CM following components have been chosen: reference current Iref = 1mA, voltage source $V_{CC} = 10V$, as a result $R = 9.3kOhm$; taking into account Early effect (assumption is that Early voltage $V_a = 100V$) output resistance is calculated to be $R_{out} =V / I = 100/1mA = 100kOhm$.
\begin{figure}  [!ht] 
\begin{center}  
\includegraphics[scale=0.5]{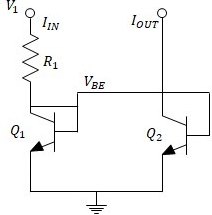} 
\caption{\small \sl Basic BJT CM schematic }
\label{noise_basic} 
\end{center}     
\end{figure}
It is known that Basic BJT CM is aimed to supply nearly constant current to a load over a wide range of load resistances. Since in LTSpice it is more convenient to change some output over varying DC voltage with specified increment in LTSpice, in order to observe mirrored current $I_{c(Q2)}$ , it has been established different loading conditions by changing load voltage VL from 0 to 10 Volts by increment of 0.1 Volt. In addition, in this case of the proposed circuit, the bipolar junction transistors have been modeled as off-the-shelf ones; because current mirrors are normally used in integrated circuits with designed for them models. The model have following parameters: a saturation current (IS) of $10^{-14}$ amperes, an area voltage (VAF) of 100 volts and current gain (BF) of transistor as 100. This model is used in all further circuits that are given in the paper.

\subsection{Widlar Current Source}
The basic current mirror can be further improved and have many variations. One of them is Widlar Current Source configuration. 
The importance of Windlar Configuration is appeared to be in a generation of a current of microampere scales without having resistors of high values at the load side \cite{Bryan}. Specifically, this is of high importance for analogue integrated circuit designing, because there have to be taken into account a large chip-surface areas of resistors with the values greater than about 50 kOhms. And Widlar in his article "Some circuit design techniques for linear integrated circuits" \cite{Widlar} proposed his circuit configuration which solves the problem that IC designers faced. The Widlar Current Mirror is represented in Figure 3.
\begin{figure}  [!ht] 
\begin{center}  
\includegraphics[scale=0.5]{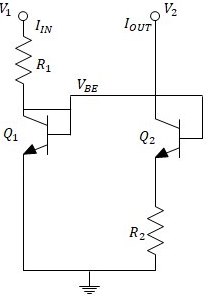} 
\caption{\small \sl Widlar Current Source schematic}
\label{noise_basic} 
\end{center}     
\end{figure}

The simulation of input and output currents against voltage of Widlar Current Mirror describes the purpose of this configuration, which is to give a small amount of current at the output, while having a high value of current at the input (Figure 4). 
\begin{figure}  [!ht] 
\begin{center}  
\includegraphics[scale=0.40]{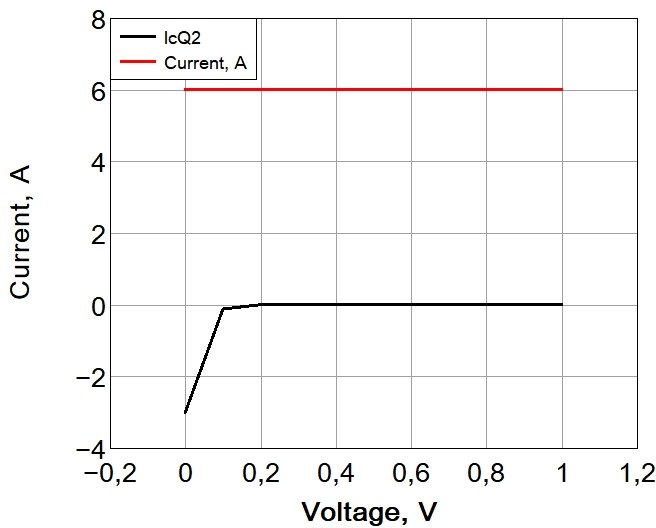} 
\caption{\small \sl Input and Output current vs. Widlar Current Source }
\label{noise_basic} 
\end{center}     
\end{figure}

\subsection{Memristor} \indent
In 1971, Professor Leon Chua, who developed memristic theory, had written an article called “Memristor-The Missing Circuit Element,” while Professor Gorm Johnsen came up with an article “An introduction to the memristor – a valuable circuit element in bioelectricity and bioimpedance.” Based on these papers, number of theoretical circuit properties, unique applications of the element were understood. Firstly, the appearance of the memristor as an missing element that characterise fundamental variables (charge and flux) in the circuit theory was introduced \cite{Leon}, \cite{bookJames}. In fact, a behaviour of the memristor is appeared to be similar to an ordinary resistor, but its resistance or conductance at particular time t is dependent on the complete past current or voltage that flowed through it \cite{Leon}, \cite{irmanova2017multi}. In addition, there are many properties that differentiates memristor from other basic circuit elements. 
\begin{figure}  [!ht] 
\begin{center}  
\includegraphics[scale=0.35]{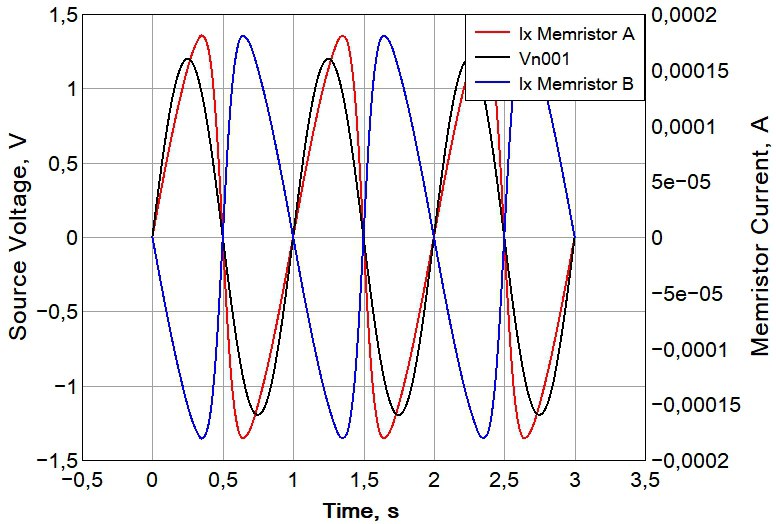} 
\caption{\small \sl Plot of Voltage V(n001) vs. Current across the memristor Ix(Memristor:A)
}
\label{noise_basic} 
\end{center}     
\end{figure}

Professor Gosm in his paper states that a memristor does not save its characteristics in charges, meaning there is no leakage of the charges, subsequently no leakage of the energy \cite{Gorm}. In other words, the data can be stored until there is an existing physical material which is made of the element itself, while other types of the storages have a tendency of aging by the time \cite{Gorm}.  The evidence for this comes from a pinched hysteresis loop in the voltage current  plane for the memristor \cite{Gorm}, \cite{irmanova2018neuron}. 
For further work it has been chosen and simulated the Joglekar Resistance Switch Memristor Model in LTSpice. As it can be seen from the figures, all of the features attributed to the memristor have been checked by simulations. 
\begin{figure}  [!ht] 
\begin{center}  
\includegraphics[scale=0.35]{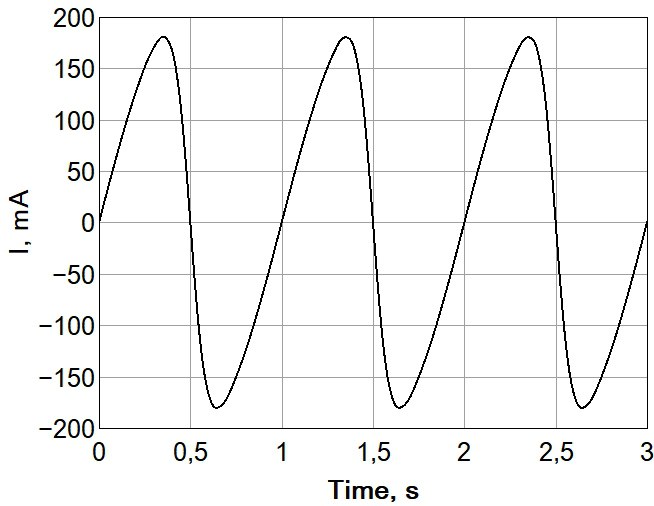} 
\caption{\small \sl Plot of source voltage V(n001), current to the input Ix(Memristor:A) and to the output Ix(Memristor:B) of the memristor
}
\label{noise_basic} 
\end{center}     
\end{figure}

\subsection{Widlar Current Mirror with memristor} \indent
	The memristor was introduced into the Widlar Current Mirror in order improve the features of the configuration. The fourth circuit element has been used instead of resistors at the input and output of the circuit as it is shown in Figure 7.
\begin{figure}  [!ht] 
\begin{center}  
\includegraphics[scale=0.4]{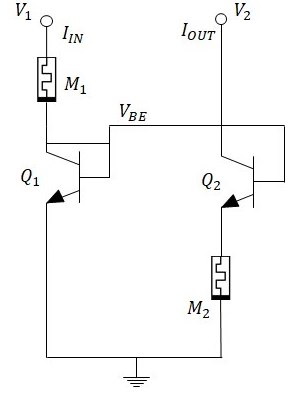} 
\caption{\small \sl Widlar Current Mirror with memristor
}
\label{noise_basic} 
\end{center}     
\end{figure}
The graph of the input and output current against load voltage in the circuit of Widlar configuration with memristor (Figure 8) have the same trend as it has been in simple Widlar Current Mirror (Figure 4). Both of the circuits state that to the given high-valued input currents and any load voltage, there is a small-valued current at the output. This proves the correctness of memristor integration into the Widlar Current Source. In the next section, the paper attempts to analyze improvements that may lead this integration.

\begin{figure}  [!ht] 
\begin{center}  
\includegraphics[scale=0.42]{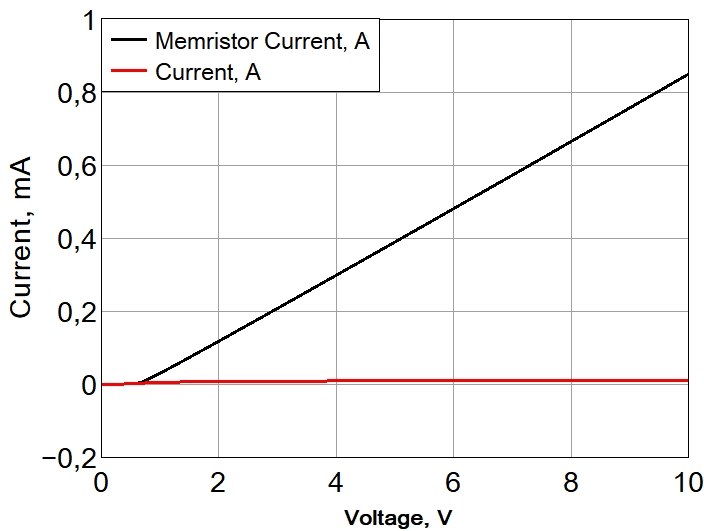} 
\caption{\small \sl Widlar Current Source with memristor:Input and Output current vs. Load Voltage}
\label{noise_basic} 
\end{center}     
\end{figure}

\section{Analysis}
\subsection{Power Analysis}
	The power analysis has been performed by calculating the power dissipation in both simple Widlar CM and improved Widlar CM with memristor. In the first case, an average power across the resistor at the input was 56.183mW, across the bjt at the input - 4.8834mW, across the resistor at the output - 362.09nW, across the bjt at the output - 128.69µW. Therefore, the total power loss of the Widlar Current Mirros was 61.2mW, which is mostly because of the input current, while the effect of the output current is minor.
    
For the Widlar Current SOurce with memristor, the dissipated power across the input memristor was 2.4124mW, across the input bjt - 296.72µW, and across the ouput memristor - 773.46nW, across the output bjt - 7.1616µW. As a result, the total power loss is 2.72mW.

Although the results has shown that a presence of memristor instead of resistor significantly decreases the power dissipation (from 61.mW to 2.72mW), it is hard to strongly conclude that this is the fact. Because the Joglekar model of memristor, that has been used in the circuit, considers an ideal parameters, which do not include the dissipation. 
\subsection{Fourier Analysis}
	Any circuit is said to operate without any trouble if the voltage or current across the circuit are at fundamental frequency. However, there are many electric elements or devices that may vary the signal waveform. Therefore, Total Harmonic Distortion (THD) is there to demonstrate the deviated rms value with respect to the fundamental frequency.
    The fourier analysis has been conducted by changing DC sources to the AC supplies with the same frequency. Then there has been written a command which calculates power of harmonics of this particular frequency. The results of the analysis can be used to compare two circuits if the resolution of simulation and number of samples are kept constant, which was done during simulation process. For the case with the simple Widlar Configuration, Total Harmonic Distortion (THD) was 1.915767\% (1.943786\%); while for the case with Widlar CM with memristor, it was  1.200688\% (1.219142\%). As a result, it can be said that THD was decreased with the memristor, meaning that there is less distortion in the circuit compared to the initial waveform of the source. 
    
\subsection{Chip-surface area analysis}
	The Widlar Current Mirror is itself, as it has been mentioned before, a solution of generating small current values at the output by avoiding a large chip-surface areas of resistors. This case is said to be improved with the integration of memristor in the circuit, because the size of the memristor, which is in nanoscales, is considerably smaller than the size of the smallest resistor necessary to generate low currents.
    
\section{Conclusion}
	To sum up, this paper is attempted, firstly, to describe the simplest current mirror, its improved version - Widlar Current Mirror, the fourth circuit element - memristor; then, to analyze the performance of the Widlar configuration with integrated memristor. The results has shown that the presence of the memristor decreases the distortion of the  signal in the circuit from the fundamental frequency waveform. In addition, the area of the chip surface is stated to be decreased. The power dissipation analysis has shown that with the memristor in the circuit the power losses may be less too. However, this not definite conclusion of this specific analysis because of the model of memristor. Nevertheless, the general performance of Widlar Current Mirror is said to be improved with the integrated memristor in the circuit. 

The further research can include comparisons of the different current mirrors with each other and with their modified versions by memristor.

\ifCLASSOPTIONcaptionsoff
  \newpage
\fi

\end{document}